\documentclass[twocolumn,showpacs,preprintnumbers,amsmath,amssymb]{revtex4}
\usepackage{mathrsfs}
\usepackage{amsfonts}

\usepackage{graphicx}
\usepackage{}
\usepackage{amsmath}
\usepackage{dcolumn}
\usepackage{color}

\begin{document}
\title{Controllable optical bistability based on photons and phonons in a two-mode optomechanical system}

\author{Cheng Jiang$^{1}$}
\email{chengjiang_8402@163.com}
\author{Hongxiang Liu$^{1,2}$}
\author{Yuanshun Cui$^{1}$}
\email{cys571015@hytc.edu.cn}
\author{Xiaowei Li$^{1}$}
\author{Guibin Chen$^{1}$}
\author{Xuemin Shuai$^{3}$}
 \affiliation{
 $^1$ School of Physics and Electronic Electrical Engineering, Huaiyin Normal University, 111 West Chang Jiang Road, Huai'an 223001, China\\
 $^2$ School of Physics, Northeast Normal University, Changchun, 130024, China\\
 $^3$ Department of physics, Chang'an University, Nan Er Huan, Xi'an, 710064, China}
\date{\today}
\begin{abstract}
We explore theoretically the bistable behavior of intracavity photon number in a two-mode cavity optomechanical system, where two cavity modes are coupled to a common mechanical resonator. When the two cavity modes are driven by two pump laser beams, respectively, we find that the optical bistability can be controlled by tuning the power and frequency of the pump beams. The common interaction to a mechanical mode enables one to control the bistable behavior in one cavity by adjusting the pump laser beam driving another cavity. We also show that both branches of optical bistability at photon numbers below unity can be observed in this two-mode optomechanical system. This phenomenon can find potential applications in controllable optical switch.
\end{abstract}

\maketitle
 \section{Introduction}
Cavity optomechanics \cite{Kippenberg2,Marquardt,Aspelmeyer} explores the interaction between a mechanical resonator and photons in a driven electromagnetic cavity via radiation pressure force. In the past decade, remarkable progress has been made in this emerging field, including quantum ground state cooling of the nanomechanical resonators \cite{Teufel2,Chan}, optomechanically induced transparency (OMIT) \cite{Agarwal,Weis,Naeini}, as well as coherent photon-phonon conversion \cite{Fiore, Verhagen, Hill} and quantum state transfer \cite{Wang, Tian, Palomaki}. The force exerted by a single photon on a macroscopic mechanical resonator is typically weak and intrinsically nonlinear. Experiments to date have focused on the regime of strong optical driving, where the optomechanical coupling can by enhanced by a factor $\sqrt{n}$, where $n$ is the mean photon number in the cavity \cite{Groblacher, Teufel}. But such enhancement comes at the expense of making the effective interaction linear. Recently, it has been reported that single-photon strong-coupling regime, where the single-photon optomechanical coupling rate $g$ exceeds the cavity decay rate $\kappa$, can be realized in single-mode \cite{Rabl, Nunenkamp, Liao, Borkje, Lemonde, Kronwald} or two-mode optomechanical systems \cite{Ludwig, Stannigel, Komar}. In this regime, the inherently nonlinear optomechanical interaction is significant at the level of single photons and phonons.

Among all the nonlinear phenomenon in a cavity optomechanical system, optical bistability is one of the focuses of research interest. Recently, the bistable behavior of the mean intracavity photon number in optomechanical systems with a Bose-Einstein condensate (BEC) \cite{Brennecke, Zhang, Yang}, ultracold atoms \cite{Gupta, Kanamoto, Purdy} and a quantum well \cite{Sete} has been extensively studied. The photon number in the optical cavity with a BEC or ultracold atoms to allow bistable behavior is usually low and even below unity due to the collective atomic motion. However, in the generic optomechanical system consisted of an empty optical cavity with one movable end mirror, optical bistability typically occurs at high photon numbers. Producing such phenomena at very low photons number is desirable for applications from optical communication to quantum computation. In the present paper, we theoretically investigate the bistable behavior of the intracavity photon number in a two-mode optomechanical system in the simultaneous presence of two strong pump laser beams and a weak probe laser beam. By adjusting the frequency and power of the pump beams, we can efficiently control the optical bistability in each cavity. In particular, one can control the bistable behavior in the right cavity by adjusting the pump beam driving the left cavity. Most importantly, such phenomena can appear at low photon numbers below unity because the two cavities are coupled to a common mechanical resonator based on photons and phonons.
The paper is organized as follows. Section \uppercase\expandafter{\romannumeral2} gives the theoretical model and method. Results and discussion are shown in Sec. \uppercase\expandafter{\romannumeral3}. A summary is presented in Sec. \uppercase\expandafter{\romannumeral4}.

\section{Model and Theory}
\begin{figure}
\includegraphics[width=8cm]{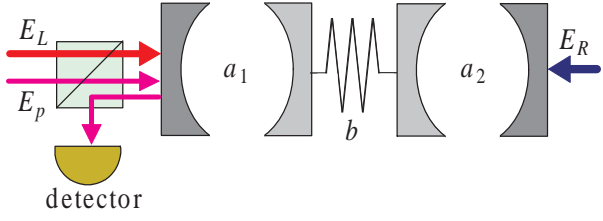}
\caption{Schematic of a two-mode optomechanical system where two optical cavity modes, $a_1$ and $a_2$, are coupled to the same mechanical mode $b$. The left cavity is driven by a strong pump beam $E_L$ in the simultaneous presence of a weak probe beam $E_p$ while the right cavity is only driven by a pump beam $E_R$.}
\end{figure}
The system under consideration is shown in Fig. 1. Two optical cavity modes $a_k$($k=1, 2$) are coupled to a common mechanical mode $b$ via an interaction Hamiltonian $H_{I}=\sum_{k=1,2}\hbar g_k
a_k^\dagger a_k(b^\dagger+b)$, where $a_k$ and $b$ are the annihilation operators of the cavity and mechanical mode, respectively, and $g_k$ is the single-photon coupling rate between the mechanical mode and the $k$th cavity mode. Physically, $g_k$ represents the frequency shift of cavity mode $k$ due to the zero-point motion of the mechanical resonator.  The left cavity is driven by a strong pump laser beam $E_L$ of frequency $\omega_L$ and a weak probe laser beam $E_p$ of frequency $\omega_p$ simultaneously, and the right cavity is only driven by a strong pump laser beam $E_R$ of frequency $\omega_R$.
 In a rotating frame at
the pump frequency $\omega_{L}$ and $\omega_R$, the Hamiltonian of the two-mode optomechanical system reads \cite{Hill}
\begin{eqnarray}
H=&&\sum_{k=1,2}\hbar\Delta_{k}a_{k}^\dagger a_{k}+\hbar\omega_{m}b^\dagger b-\sum_{k=1,2}\hbar g_k
a_k^\dagger a_k(b^\dagger+b)\nonumber\\&&+i\hbar\sqrt{\kappa_{e,1}}E_{L}(a_{1}^\dagger -a_{1})+i\hbar\sqrt{\kappa_{e,2}}E_{R}(a_{2}^\dagger -a_{2})\nonumber\\&&+i\hbar\sqrt{\kappa_{e,1}}E_{p}(a_{1}^\dagger
e^{-i\delta t}-a_{1}e^{i\delta t}).
\end{eqnarray}
The first term represents the energy of the two optical cavity modes with resonance frequency $\omega_k(k=1,2)$, where
$\Delta_1=\omega_1-\omega_L$ and $\Delta_2=\omega_2-\omega_R$ are the corresponding cavity-pump
field detunings. The second term gives the energy of the
mechanical mode resonance frequency $\omega_{m}$ and effective
mass $m$. The last three terms describe the interaction between the
input fields and the cavity fields, where $E_L$, $E_R$, and $E_p$ are related to the power of the applied laser fields by $\left\vert
E_L\right\vert=\sqrt{2P_L\kappa_1/\hbar\omega_L}$, $\left\vert
E_R\right\vert=\sqrt{2P_R\kappa_2/\hbar\omega_R}$, and $\left\vert
E_p\right\vert=\sqrt{2P_p\kappa_1/\hbar\omega_p}$ ($\kappa_k$ the linewidth of the \emph{k}th cavity mode), respectively. Each optical cavity is coupled not only to a shared mechanical mode, but also to an optical bath at rate $\kappa_{i,k}$ and to an external photonic waveguide at rate $\kappa_{e,k}$. Therefore, the total cavity linewidth $\kappa_k=\kappa_{i,k}+\kappa_{e,k}$.
Here, $\delta=\omega_p-\omega_L$ is the detuning between the probe laser beam and the left pump laser beam.

According to the Heisenberg equations of motion and the commutation relation $[a_k,a_k^\dagger]=1$, and $[b,b^\dagger]=1$, the temporal evolutions of operators $a_1$, $a_2$, and $Q$ [which is defined as $Q=b^\dagger+b$] can be obtained. Introducing the corresponding damping and noise terms for the mechanical and cavity modes, we derive the quantum Langevin equations as follows:
\begin{eqnarray}
&&\dot{a_1}=-i(\Delta_1-g_1Q)a_1-\kappa_1 a_1+\sqrt{\kappa_{e,1}}(E_L+E_pe^{-i\delta t})\nonumber\\&&+\sqrt{2\kappa_1}a_{in,1},\\
&&\dot{a_2}=-i(\Delta_2-g_2Q)a_2-\kappa_2 a_2+\sqrt{\kappa_{e,2}}E_R\nonumber\\&&+\sqrt{2\kappa_2}a_{in,2},\\
&&\ddot{Q}+\gamma_m\dot{Q}+\omega_m^2 Q=2g_1\omega_ma_1^\dagger a_1+2g_2\omega_m a_2^\dagger a_2+\xi.
\end{eqnarray}
The cavity modes decay at the rate $\kappa_k (k=1,2)$ and are affected by the input vacuum noise operator $a_{in,k}$ with zero mean value, which obey the correlation function in the time domain \cite{Genes},
\begin{eqnarray}
&\langle\delta a_{in,k}(t)\delta a_{in,k}^\dagger(t^\prime)\rangle=\delta(t-t^\prime),\\
&\langle\delta a_{in,k}(t)\delta a_{in,k}(t^\prime)\rangle=\langle\delta a_{in,k}^\dagger(t)\delta a_{in,k}(t^\prime)\rangle=0,
\end{eqnarray}
The mechanical mode is affected by a vicious force with damping rate $\gamma_m$ and by a Brownian stochastic force with zero mean value $\xi$ that has the following correlation function \cite{Giovannetti}
\begin{eqnarray}
\langle\xi(t)\xi(t^\prime)\rangle=\frac{\gamma_m}{\omega_m}\int \frac{d\omega}{2\pi}\omega e^{-i\omega (t-t^\prime)}\left[1+\mathrm{coth}\left(\frac{\hbar\omega}{2k_BT}\right)\right],
\end{eqnarray}
where $k_B$ is the Boltzmann constant and $T$ is the temperature of the reservoir of the mechanical resonator.

Setting all the time derivatives to zero, we derive the steady-state solution to Eqs. (2)-(4)
\begin{eqnarray}
&&a_{s,1}=\frac{\sqrt{\kappa_{e,1}}E_L}{\kappa_1+i\Delta_1'}, a_{s,2}=\frac{\sqrt{\kappa_{e,2}}E_R}{\kappa_2+i\Delta_2'},\nonumber\\&&
Q_s=\frac{2}{\omega_m}(g_1\left\vert a_{s,1}\right\vert^2+g_2\left\vert a_{s,2}\right\vert^2),
\end{eqnarray}
where $\Delta_1'=\Delta_1-g_1 Q_s$ and $\Delta_2'=\Delta_2-g_2 Q_s$ are the effective cavity detunings including radiation pressure effects. Therefore, the mean intracavity photon numbers $n_{pk}=\left\vert a_{s,k}\right\vert^2$ can be determined by the following coupled equations
\begin{eqnarray}
n_{p1}=\frac{\kappa_{e,1}E_L^2}{\kappa_1^2+\left[\Delta_1-2g_1/\omega_m(g_1n_{1}-g_2n_{2})\right]^2},\\
n_{p2}=\frac{\kappa_{e,2}E_R^2}{\kappa_2^2+\left[\Delta_2-2g_2/\omega_m(g_1n_{1}-g_2n_{2})\right]^2}.
\end{eqnarray}
This form of coupled cubic equations are characteristic of the optical multistability \cite{Gupta, Kanamoto}. Recently, optical bistability of mean intracavity photon numbers in optomechanical systems with ultracold atoms \cite{Gupta} and a BEC \cite{Ritter} have been experimentally observed in the transmission of the probe laser beam through the cavity. Here we theoretically propose a flexible scheme to control the bistable behavior in the two-mode optomechanical system according to Eqs. (9)-(10). In essence it is the nonlinearity owing to optomechanical coupling that is responsible for the bistable behavior in this scheme.

\section{Numerical results and discussion}
We consider for illustration an experimentally realized two-mode optomechanical system. The parameters used are \cite{Hill}: $\omega_1=2\pi\times205.3$ THz, $\omega_2=2\pi\times194.1$ THz, $\kappa_1=2\pi\times520$ MHz, $\kappa_2=1.73$ GHz, $\kappa_{e,1}=0.2\kappa_1$, $\kappa_{e,2}=0.42\kappa_2$, $g_1=2\pi\times960$ kHz, $g_2=2\pi\times430$ kHz, $\omega_m=2\pi\times4$ GHz, $Q_m=87\times10^3$,
 where $Q_{m}$ is the quality
factor of the nanomechanical resonator, and the damping rate $\gamma
_{m}$ is given by $\frac{\omega _{m}}{Q_m}.$ We can see that
$\omega_{1}>\kappa_1$ and $\omega_2>\kappa_2$, therefore the system would be in the good-cavity limit, a prerequisite for ground-state cooling of nanomechanical resonators \cite{Teufel2}.

\begin{figure}
\includegraphics[width=8cm]{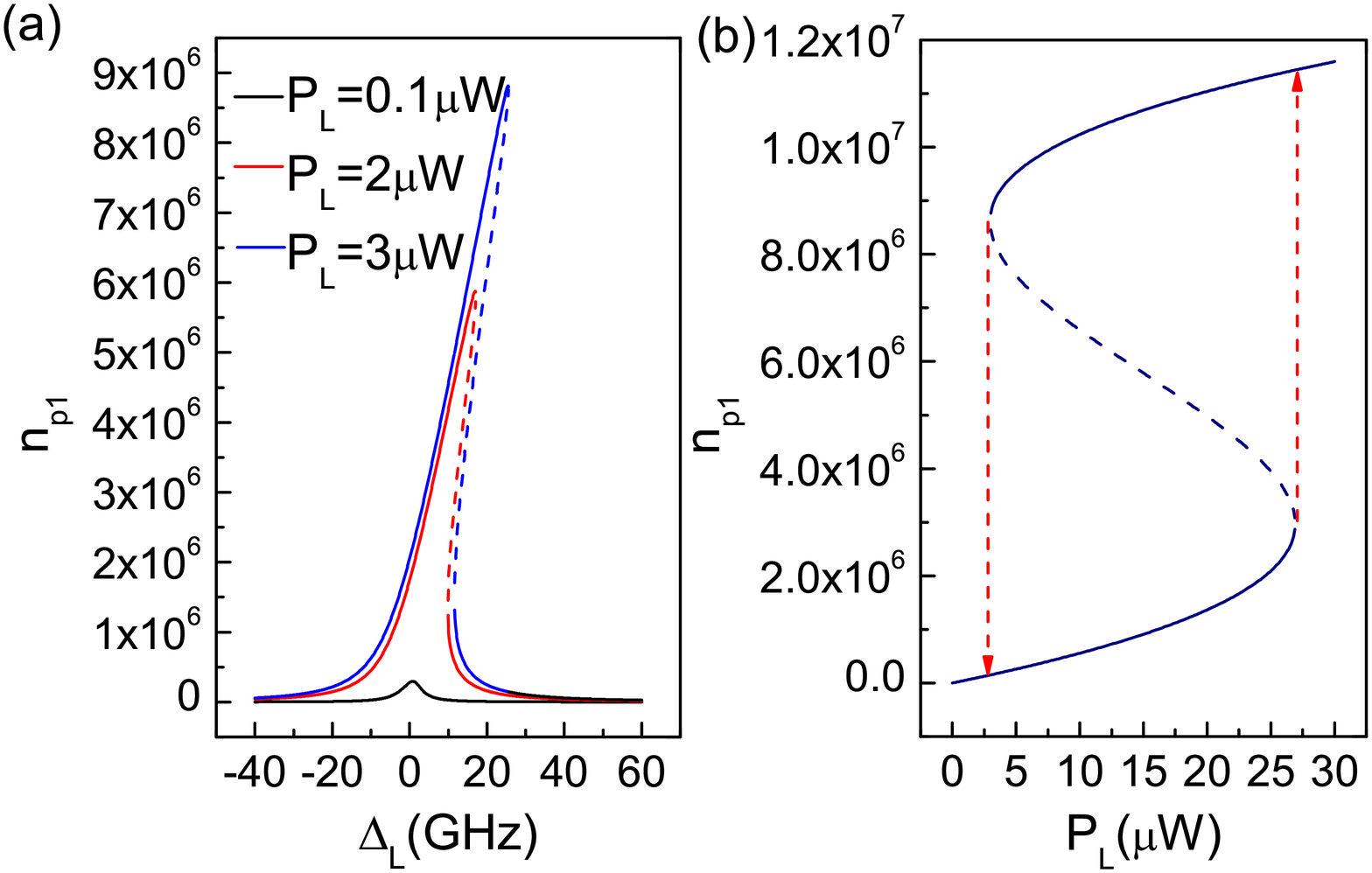}
\caption{Mean intracavity photon number of the left cavity as a function of (a) the cavity-pump detuning $\Delta_L=\omega_1-\omega_L$ for left pump power
$P_L$ equals to 0.1, 2, and 3 $\mu$W(from bottom to top) (b) the left pump power $P_L$ for $\Delta_L=\Delta_R=\omega_m.$ The right pump power is kept equal to 0.1 $\mu$W.
Other parameters used are $\omega_1=2\pi\times205.3$ THz, $\omega_2=2\pi\times194.1$ THz, $\kappa_1=2\pi\times520$ MHz, $\kappa_2=1.73$ GHz, $\kappa_{e,1}=0.2\kappa_1$, $\kappa_{e,2}=0.42\kappa_2$, $\omega_m=2\pi\times4$ GHz, $Q_m=87\times10^3$, $g_1=2\pi\times960$ kHz, and $g_2=2\pi\times430$ kHz.}
\end{figure}
The two-mode optomechanical system we consider here enables more controllability in the bistable behavior of the intracavity photon number. Characterization of the optomechanical cavity can be performed by using two strong pump beams combined with a weak probe beam. Here, we mainly investigate the optical bistability in each cavity by adjusting the power and frequency of the left pump beam. Figure 2(a) plots the mean intracavity photon number in the left cavity as a function of the left cavity-pump detuning $\Delta_L=\omega_1-\omega_L$ for various pump powers. When the power of the the left pump beam is $P_L=0.1\mu$W, the curve is nearly Lorentzian. However, when the power increases above a critical value, the system exhibits bistable behavior as shown in the curves for $P_L=2\mu$W and $P_L=3\mu$W, where the initially Lorentzian resonance curve becomes asymmetric. In this case, the coupled cubic equations (9)-(10) for the mean intracavity photon number yields three real roots. The largest and smallest roots are stable, and the middle one is unstable, which is represented by the dashed lines in Fig. 2(a). Furthermore, we can see that larger cavity-pump detuning is necessary to observe the optical bistable behavior with the increasing pump beam power. The bistable behavior can also be seen from the hysteresis loop for the mean intracavity photon number versus the pump power curve shown in Fig. 2(b). Here, both the cavities are pumped on their respective red sidebands, i.e., $\Delta_L=\Delta_R=\omega_m$, which is beneficial for resolved sideband cooling of the nanomechanical resonator \cite{Teufel2}. Consider the left pump power increases from zero gradually, the mean intracavity photon number $n_{p1}$ initially lies in the lower stable branch (corresponding to the smallest root). When the pump power $P_L$ increases to a critical value, about $27\mu$W in our case, $n_{p1}$ approaches the end of this branch. The hysteresis then follows the arrow and jumps to the upper branch. If the $P_L$ is increased further, $n_{p1}$ remains on the upper branch. If $P_L$ is decreased, $n_{p1}$ gradually approaches the beginning of the second stable branch. If $P_L$ is decreased even further, the hysteresis follows the arrow and switches back to the lower stable branch. Based on the above discussions, one can easily realize an optical switch by modulating the power and frequency of the pump beam.
\begin{figure}
\centering
\includegraphics[width=8cm]{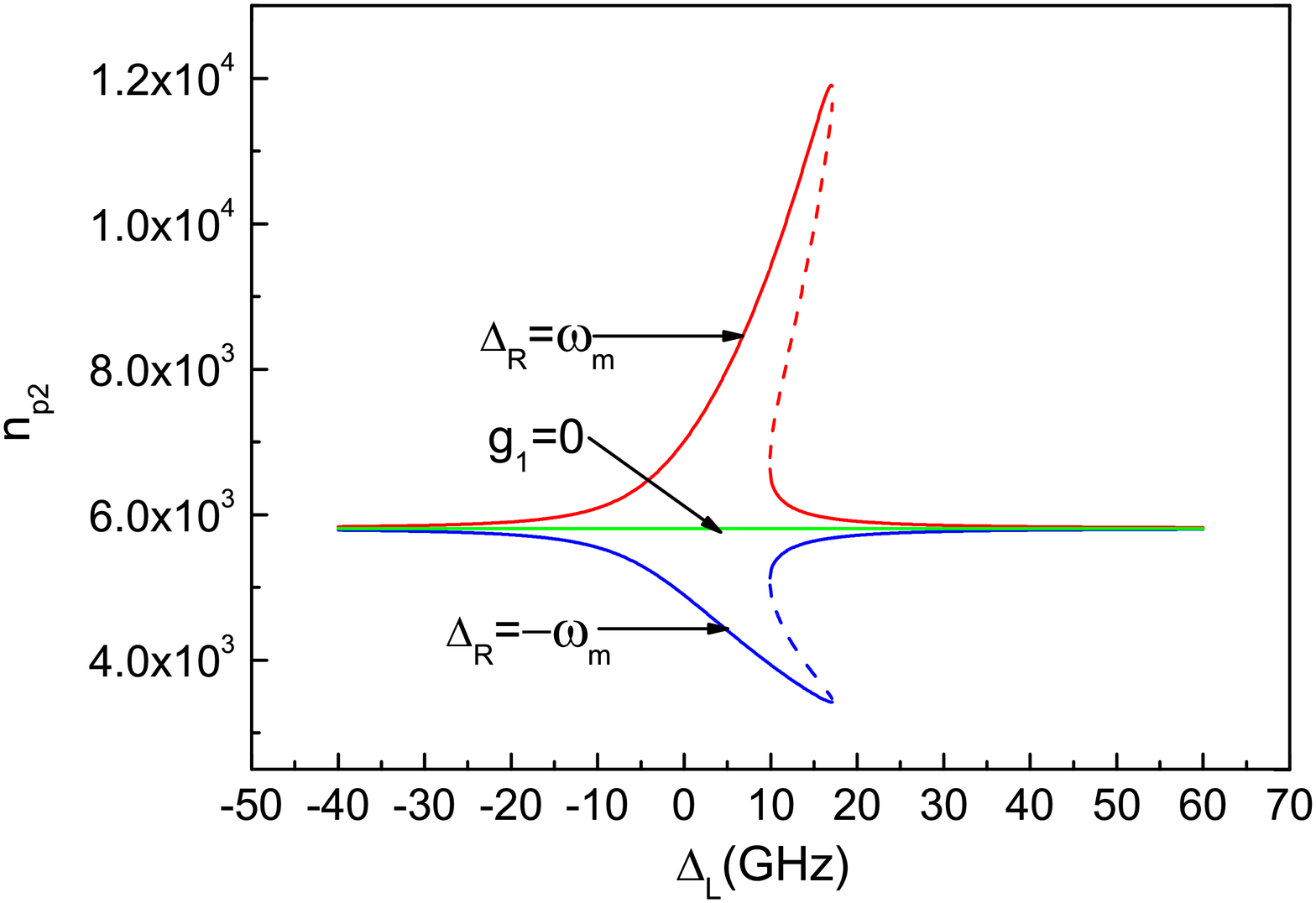}
\caption{Mean intracavity photon number of the right cavity versus the left cavity-pump detuning $\Delta_L$ with $\Delta_R=\omega_m$ and $\Delta_R=-\omega_m$, respectively. The left pump power $P_L$ equals to $2\mu W$ and the right pump power $P_R$ equals to $0.1\mu W$. When the coupling strength between the left cavity and the mechanical resonator $g_1$ is equal to zero and $\Delta_R=\omega_m$, the intracavity photon number keeps constant. Other parameters are the same as in figure 2.}
\end{figure}

In the following, we mainly investigate the optical bistability in the right cavity by controlling the frequency and power of the left pump beam. Mean intracavity photon number $n_{p2}$ in the right cavity as a function of the left cavity-pump beam detuning $\Delta_L$ is plotted in Fig. 3. The left pump power $P_L=2\mu$W and the right pump power $P_R=0.1\mu$W. When the coupling between the left cavity and the mechanical resonator turns off, i.e., $g_1=0$, the two-mode optomechanical system becomes the generic single-mode optomechanics, and the pump beam driving the left cavity can't have an impact on the photon numbers in the right cavity via the mechanical mode. In this case, if the the right cavity is pumped on its red sideband ($\Delta_R=\omega_m$), it can be seen clearly from the middle straight line in Fig. 3 that the mean intracavity photon number $n_{p2}$ keeps constant when the left cavity-pump detuning $\Delta_L$ changes. However, if the coupling between the left cavity and the nanomechanical resonator turns on, bistable behavior of the mean intracavity photon number in the right cavity will appear. When $\Delta_R=\omega_m$, the average photon number is larger than the constant value obtained before. However, if the right cavity is pumped on its blue sideband, i.e., $\Delta_R=-\omega_m$, the average photon number is smaller than the above constant value. The underling physical mechanism for this phenomenon can be explained as follows. When $g_1=0$ and $\Delta_R=\omega_m$, the hybrid system turns to the typical single mode optomechanical system, and the intracavity photon number is directly related to the pump power $P_R$ and the square of the cavity-pump detuning (${\Delta_R}^2$) \cite{Teufel}. \begin{figure}
\centering
\includegraphics[width=8cm]{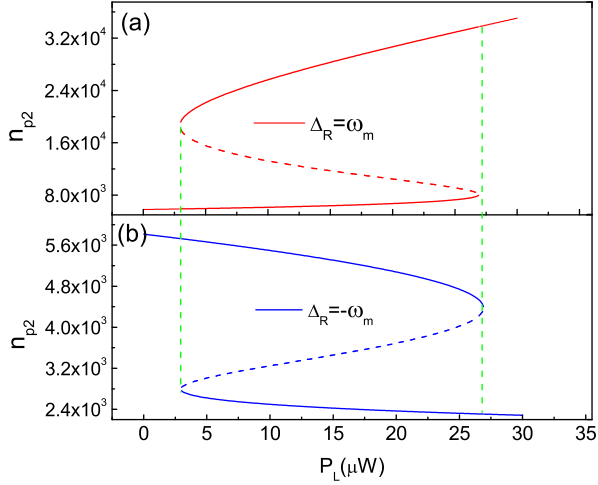}
\caption{Mean intracavity photon number of the right cavity as a function of the left pump power for $P_R=0.1\mu$W with (a) $\Delta_R=\omega_m$ (b) $\Delta_R=-\omega_m$, respectively. The left cavity-pump detuning $\Delta_L$ is kept equal to $\omega_m$. Other parameters are the same as in figure 2.}
\end{figure}Therefore, the photon number keeps constant when $g_1=0,P_R=0.1\mu$W, and $\Delta_R=\pm\omega_m$. However, when $g_1\neq0$, the intracavity photons in the left cavity will have an effect on the common nanomechanical resonator and subsequently the photon number in the right cavity. The simultaneous presence of the left pump and probe beam induces a radiation pressure force at the beat frequency $\delta=\omega_p-\omega_L$, which drives the mechanical mode to oscillate coherently. When $\Delta_R=\omega_2-\omega_R=\omega_m$, the highly off-resonant Stokes scattering at the frequency $\omega_R-\omega_m$ is strongly suppressed and only the anti-Stokes scattering at the frequency $\omega_R+\omega_m$ builds up within the right cavity, leading to the up-conversion of the pump photons to the cavity photons at the frequency $\omega_2$. Therefore, the average photon number in the right cavity is lager than the constant value without of the effect of the left cavity. On the other hand, when $\Delta_R=-\omega_m$, the anti-Stokes scattering is strongly suppressed while the Stokes scattering at the frequency $\omega_R-\omega_m$ builds up within the cavity, leading to the down-conversion of cavity photons at the frequency $\omega_2$ to the pump photons. In this case, the average photon number in the right cavity is smaller than the constant value when $g_1=0$. Therefore, by adjusting the left cavity-pump beam detuning $\Delta_L$, one can observe the bistable behavior of the intracavity photon number in the right cavity.

Optical bistability in the right cavity can also be seen from the hysteresis loop for the mean intracavity photon number versus the left pump power when $\Delta_R=\omega_m$ and $\Delta_R=-\omega_m$, as shown in Fig. 4. Here we have taken left cavity-pump beam detuning to be $\Delta_L=\omega_m$ and $P_R=0.1\mu$W. Similarly, mean intracavity photon number when $\Delta_R=\omega_m$ is larger than the situation when $\Delta_R=-\omega_m$. Different from the curve for $\Delta_R=\omega_m$, $n_{p2}$ initially lies in the upper stable branch (corresponding to the largest root) if $\Delta_R=-\omega_m$. With the increase of the pump power $P_L$, $n_{p2}$ decreases gradually and jumps to the lower branch when $P_L$ approaches a critical value, about $27\mu$W in our case. We can also realize a switch between the lower and upper stable branch by controlling the power of the left pump beam.
 
\begin{figure}
\centering
\includegraphics[width=8cm]{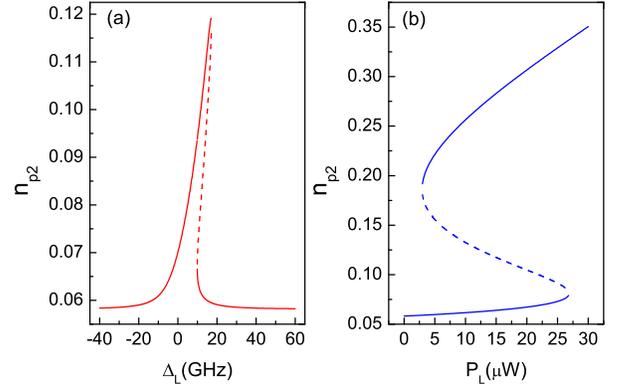}
\caption{Mean intracavity photon number of the right cavity versus (a) the left cavity-pump detuning $\Delta_L$ for $P_L=2\mu$W (b) the left pump power $P_L$ for $\Delta_L=\omega_m$. Other parameters used are $P_R=1$pW, $\Delta_R=\omega_m$, $\omega_1=2\pi\times205.3$ THz, $\omega_2=2\pi\times194.1$ THz, $\kappa_1=2\pi\times520$ MHz, $\kappa_2=1.73$ GHz, $\kappa_{e,1}=0.2\kappa_1$, $\kappa_{e,2}=0.42\kappa_2$, $\omega_m=2\pi\times4$ GHz, $Q_m=87\times10^3$, $g_1=2\pi\times960$ kHz, and $g_2=2\pi\times430$ kHz.}
\end{figure}
In our previous discussions, we have demonstrated optical bistability in both cavities, and the mean intracavity photon number is usually very large, at least thousands of photons in the right cavity (see Fig. 3 and 4). For the single-mode optomechanical system ($g_2=0$), it would need much more photons in the cavity in order to reach the bistable regime (see Fig. 2 for illustration) \cite{Sete}. In what follows, we will show that two-mode optomechanical system allows for optical bistability at extremely low cavity photon numbers. Fig. 5 plots the mean photon number in the right cavity as a function of (a) left cavity-pump beam detuning $\Delta_L$ for $P_L=2\mu$W and (b) left pump power $P_L$ for $\Delta_L=\omega_m$. Parameters of the right pump beam are $P_R=1$pW and $\Delta_R=\omega_m$. Due to the low pump power, intracavity photon number in the right cavity is very small, i.e., $n_{p2}\leq1$. Generally, such low photon numbers cannot exhibit bistable behavior in the empty cavity optomechanical systems. However, in the two-mode optomechanics we consider here, when the left cavity is driven by a strong pump power, because the two cavities are coupled to a common nanomechanical resonator, optical bistability can still exist in the right cavity at the extremely low intracavity photon numbers. This phenomenon signifies the strong nonlinear effects in the weak coupling regime \cite{Gupta}, which is enabled by the long lifetime of the mechanical mode and the strong pump on the left cavity. Recently, two related works by L\"{u} \emph{et al.} \cite{Lu} and Kuzyk \emph{et al.} \cite{Kuzyk} have also shown that strong nonlinearities can be obtained in two-mode optomechanical systems in the weak coupling regime. As pointed out by previous works, the bistable behavior of intracavity photon numbers in the two-mode optomechanical system under consideration also provides a candidate of realizing a controllable optical switch. For this, the two stable branches of photon numbers in the right cavity act as the optical switch. When the frequency and power of the left pump beam are fixed, the switch between the lower stable branch and the upper stable branch can be easily realized by controlling the frequency and power of the right pump beam. Furthermore, the left pump beam can be used as a control parameter to enable or disable this switch. When the photon number in the right cavity is less than one, if we turn off the left pump beam, the bistable behavior disappears and only one stable branch exists.

\section{conclusion}
In conclusion, we have investigated the optical bistability in a two-mode optomechanical system where two optical cavities are coupled to a common mechanical resonator via radiation pressure force. Compared with the generic single-mode cavity optomechanics, such a two-mode optomechanical system allows one to control the optical bistability in a much more flexible way. The bistable behavior of the mean intracavity photon number in one cavity can be tuned by the power and frequency of the pump laser beam driving another cavity. Furthermore, bistability at low photon numbers below unity should be possible in such a coupled system.

\section{acknowledgments} The authors gratefully acknowledge support
from National Natural Science Foundation of China (Grant Nos.11074088 and 11174101) and Jiangsu Natural Science Foundation (Grant No. BK2011411).


\begin{thebibliography}{32}
\bibitem{Kippenberg2} T. J. Kippenberg and K. J. Vahala, Science \textbf{321,} 1172 (2008).

\bibitem{Marquardt} F. Marquardt and S. M. Girvin, Physics \textbf{2,} 40 (2009).

\bibitem{Aspelmeyer} M. Aspelmeyer, T. J. Kippenberg, F. Marquardt, arXiv:1303.0733.

\bibitem{Teufel2}J. D. Teufel, T. Donner, D. Li, J. W. Harlow, M. S. Allman, K. Cicak, A. J. Sirois, J. D. Whittaker, K. W. Lehnert, and R. W. Simmonds, Nature (London) \textbf{475,} 359 (2011).

\bibitem{Chan}J. Chan, T. P. Alegre, A. H. Safavi-Naeini, J. T. Hill, A. Krause, S. Groeblacher, M. Aspelmeyer, and O. Painter, Nature (London) \textbf{478,} 89 (2011).

\bibitem{Agarwal} G. S. Agarwal and S. Huang, Phys. Rev. A \textbf{81}, 041803
(2010).

\bibitem{Weis} S. Weis, R. Rivi\`{e}re , S. Del\'{e}glise, E. Gavartin,
O. Arcizet, A. Schliesser, and T. J. Kippenberg, Science
\textbf{330}, 1520 (2010).

\bibitem{Naeini} A. H. Safavi-Naeini, T. P. Mayer Alegre, J. Chan, M. Eichenfield, M. Winger, Q. Lin, J. T.Hill, D.
E. Chang, and O. Painter, Nature (London) \textbf{472}, 69 (2011).

\bibitem{Fiore}V. Fiore, Y. Yang, M. C. Kuzyk, R. Barbour, L. Tian, and H. Wang, Phys. Rev. Lett. \textbf{107,} 133601 (2011).

\bibitem{Verhagen}E. Verhagen, S. Del\'{e}glise, S. Weis, A. Schliesser, and T. J. Kippenberg, Nature (London) \textbf{482,} 63 (2012).

\bibitem{Hill} J. T. Hill, A. H. Safavi-Naeini, J. Chan, O. Painter, Nat. Commun. \textbf{3,} 1196 (2012).

\bibitem{Wang} Y. D. Wang and A. A. Clerk, Phys. Rev. Lett. \textbf{108,} 153603 (2012).

\bibitem{Tian} L. Tian, Phys. Rev. Lett. \textbf{108,} 153604 (2012).

\bibitem{Palomaki} T. A. Palomaki, J. W. Harlow, J. D. Teufel, R. W. Simmonds, and K. W. Lehnert, Nature (London) \textbf{495,} 210 (2013).

\bibitem{Groblacher}S. Gr\"{o}blacher, K. Hammerer, M. R. Vanner, and M.
Aspelmeyer, Nature (London) \textbf{460,} 724 (2009).

\bibitem{Teufel}J. D. Teufel, D. Li, M. S. Allman,
K. Cicak, A. J. Sirois, J. D. Whittaker, and R. W. Simmonds, Nature (London)
\textbf{471}, 204 (2011).

\bibitem{Rabl}P. Rabl, Phys. Rev. Lett. \textbf{107,} 063601 (2011).

\bibitem{Nunenkamp}A. Nunnenkamp, K. B{\o}rkje, and S. M. Girvin, Phys. Rev. Lett. \textbf{107,} 063602 (2011).

\bibitem{Liao} J. Q. Liao, H. K. Cheung, and C. K. Law, Phys. Rev. A \textbf{85,} 025803 (2012).

\bibitem{Borkje} K. B{\o}rkje, A. Nunnenkamp, J. D. Teufel, S. M. Girvin, Phys. Rev. Lett. \textbf{111,} 053603 (2013).

\bibitem{Lemonde} M. A. Lemonde, N. Didier, and A. A. Clerk, Phys. Rev. Lett. \textbf{111,} 053602 (2013).

\bibitem{Kronwald} A. Kronwald and F. Marquardt, arXiv:1304.5230.

\bibitem{Ludwig}M. Ludwig, A. H. Safavi-Naeini, O. Painter, and F. Marquardt, Phys. Rev. Lett. \textbf{109,} 063601 (2012).

\bibitem{Stannigel}K. Stannigel, P. Komar, S. J. M. Habraken, S. D. Bennett, M. D. Lukin, P. Zoller, and P. Rabl, Phys. Rev. Lett. \textbf{109,} 013603 (2012).

\bibitem{Komar}P. K\'{o}m\'{a}r, S. D. Bennett, K. Stannigel, S. J. M. Habraken, P. Rabl, P. Zoller, and M. D. Lukin, Phys. Rev. A \textbf{87,} 013839 (2013).

\bibitem{Brennecke} F. Brennecke, S. Ritter, T. Donner, and T. Esslinger, Science \textbf{322,} 235 (2008).

\bibitem{Zhang} J. M. Zhang, F. C. Cui, D. L. Zhou, and W. M. Liu, Phys. Rev. A \textbf{79,} 033401 (2009).

\bibitem{Yang} S. Yang, M. Amri, and M. S. Zubairy, Phys. Rev. A \textbf{87,} 033836 (2013).

\bibitem{Gupta} S. Gupta, K. L. Moore, K. W. Murch, and D. M. Kurn, Phys. Rev. Lett. \textbf{99,} 213601 (2007).

\bibitem{Kanamoto} R. Kanamoto and P. Meystre, Phys. Rev. Lett. \textbf{104,} 063601 (2010).

\bibitem{Purdy} T. P. Purdy, D. W. C. Brooks, T. Botter, N. Brahms, Z.-Y. Ma, and D. M. Stamper-Kurn, Phys. Rev. Lett. \textbf{105,} 133602 (2010).

\bibitem{Sete} E. A. Sete and H. Eleuch, Phys. Rev. A \textbf{85,} 043824 (2012).

\bibitem{Genes} C. Genes, D. Vitali, P. Tombesi, S. Gigan, and M. Aspelmeyer, Phys. Rev. A \textbf{77,} 033804 (2008).

\bibitem{Giovannetti} V. Giovannetti and D. Vitali, Phys. Rev. A \textbf{63,} 023812 (2001).

\bibitem{Ritter} S. Ritter, F. Brennecke, K. Baumann, T. Donner, C. Guerlin, and T. Esslinger, Appl. Phys. B \textbf{95,} 213 (2009).

\bibitem{Lu} X. Y. L\"{u}, W. M. Zhang, S. Ashhab, Y. Wu, F. Nori, arXiv:1210.8299.

\bibitem{Kuzyk} M. C. Kuzyk, S. J. Enk, H. L. Wang, arXiv:1307.2844.

\end{thebibliography}
\end{document}